\documentclass[12pt,a4paper]{article}

\usepackage{amsmath}
\usepackage{amsthm}
\usepackage{amsfonts}
\usepackage{artA4}
\usepackage[T2A]{fontenc}
\usepackage[cp1251]{inputenc}
\usepackage{graphicx}
\usepackage{wrapfig}

\newtheorem{pro}{Proposition}
\newtheorem{teo}{Theorem}
\theoremstyle{remark}
\newtheorem{rem}{Remark}

\begin{document}
\renewcommand{\figurename}{Fig}
\renewcommand{\abstractname}{Abstract}
\renewcommand{\refname}{References}
\renewcommand{\refname}{References}

\title{Hamiltonization of nonholonomic systems}

\author{A.\,V.\,Borisov, I.\,S.\,Mamaev}
\date{}
\maketitle

\begin{abstract}
We consider some issues of the representation in the Hamiltonian form of two problems of
nonholonomic mechanics, namely, the Chaplygin's ball problem and the
Veselova problem. We show that these systems can be written as
generalized Chaplygin systems and can be integrated by the method
of reducing multiplier. We also indicate the algebraic form of the Poisson
brackets of these systems (after the time substitution).
Generalizations of the problems are considered
and new realizations of
nonholonomic constraints are presented. Some nonholonomic systems with
an invariant measure and a sufficient number of first integrals are indicated, for which
 the question of the representation in the Hamiltonian form is still open, even after the
time substitution.
\end{abstract}

First, we consider certain general results related to the method of
integration of nonholonomic systems, which S.\,A.\,Chaplygin \cite{b003}
called the method of reducing multiplier. We generalize this method so it can
apply to a broader class of systems, so-called generalized Chaplygin
systems. In the remaining sections, we apply these results to find
explicitly the Poisson structures and the isomorphisms with other
integrable Hamiltonian systems.

\section{Generalized Chaplygin systems}\label{sec2904-1}

Consider a mechanical system with two degrees of freedom, such that
the equations of motion can be written in the form:
\begin{equation}
\label{eq2704-7}
\begin{gathered}
\frac d{dt}\biggl(\frac{\partial L}{\partial\dot q_1}\biggr)-
\frac{\partial L}{\partial q_1}=\dot{q}_2S,\quad\frac d{dt}\biggl(
\frac{\partial L}{\partial\dot{q}_2}\biggr)-\frac{\partial L}{\partial q_2}=-\dot{q}_1S,\\
S=a_1(q)\dot{q}_1+a_2(q)\dot{q}_2+b(q),
\end{gathered}
\end{equation}
where $L$ is a function of coordinates and velocities; this function
will be also referred to as the Lagrangian of the system.

For a special form of $S$ with $b(q)=0$ we obtain an ordinary Chaplygin system \cite{b003}.
S.\,A.\,Chaplygin showed that the equations of so-called Chaplygin sleigh can be reduced to the form
\eqref{eq2704-7} (with $b(q)=0$); the system \eqref{eq2704-7} can be integrated
by applying the below-stated
method of reducing multiplier and the solution of the Hamilton--Jacobi
equation. In \cite{b001} the authors show that the Veselova system is
a Chaplygin system. The Veselova system describes the rotation of a rigid body about a
fixed point under a nonintegrable constraint: the projection of its
angular velocity onto the fixed axis is zero. We show below that
the system of equations for Chaplygin's
ball on a plane \cite{b006} can also be reduced to the form \eqref{eq2704-7} (but
with $b(q)\neq0$).

We will call the system \eqref{eq2704-7} {\it a generalized Chaplygin
system} (it should not be confused with that from \cite{b022, koil}, where a
different generalization of Chaplygin systems was offered!).

Chaplygin showed that in the case $b(q)=0$ \cite{b003}, the equations
preserve their form under time substitutions
$$
N(q)\,dt=d\tau
$$
if $N$ does not depend on the velocities. Let us show that this also holds
for equations \eqref{eq2704-7}.

Denoting the differentiation by
$q_i'=\displaystyle\frac{dq_i}{d\tau}$ we find
$$
\dot q_i=Nq_i',\quad\frac{\partial L}{\partial\dot q_i}=
\frac1N\frac{\partial\bar L}{\partial q_i'},\quad\frac{\partial L}{\partial q_i}=
\frac{\partial\bar L}{\partial q_i}-\frac1N\frac{\partial N}{\partial q_i}
\sum\limits_{k=1}^2q_k'\frac{\partial\bar L}{\partial q_k'},
$$
where $\bar L(q,q')=L(q,Nq')$.

Substituting into \eqref{eq2704-7} gives
\begin{equation}
\label{eq2704-8}
\begin{gathered}
\frac d{d\tau}\left(\frac{\partial\bar L}{\partial q_q'}\right)-
\frac{\partial\bar L}{\partial q_1}=q_2'\bar S,\quad\frac d{d\tau}
\left(\frac{\partial\bar L}{\partial q_2'}\right)-
\frac{\partial\bar L}{\partial q_2}=-q_1'\bar S,\\
\bar S=NS+\frac1N\left(\frac{\partial N}{\partial q_2}\frac{\partial\bar L}{\partial q_1'}-
\frac{\partial N}{\partial q_1}\frac{\partial\bar L}{\partial q_2'}\right).
\end{gathered}
\end{equation}

It is known for an ordinary Chaplygin system \cite{b003} that if there is
an invariant measure with density depending only on the coordinates, one
can choose $N(q)$ for which $\bar S=0$; hence, in terms of the new time
$\tau$, the system can be written in the classical Hamiltonian form. Let
us give a generalization of this result for the case of generalized
Chaplygin systems of the form \eqref{eq2704-7}, under the assumption that the
Lagrangian is a quadratic function of the velocities $\dot q_i$ (not
necessarily homogeneous).

\begin{teo}
\label{teo2904-1} Let
$\displaystyle\det\left\|\frac{\partial^2L}{\partial\dot q_i\partial\dot
q_j}\right\|\neq0$, and let the system \eqref{eq2704-7} admit an
invariant measure with density depending only on the coordinates; then
there is a time substitution $N(q)\,dt=d\tau$ such that

$1)$ the function $\bar S$, defined by \eqref{eq2704-8}, depends only on
the coordinates\/{\rm:} $\bar S=\bar S(q)$,

$2)$ in terms of the new time,
the equations of motion can be written in Hamiltonian form:
$$
\frac{dq_i}{d\tau}=\{q_i,\bar H\},\quad\frac{dp_i}{d\tau}=\{p_i,\bar H\},
$$
where
$$
p_i=\frac{\partial\bar L}{\partial q_i'},\quad\bar H=\sum\limits_{k=1}^2p_kq_k'-
\bar L\Bigr|_{q_i'\to p_i},
$$
and the Poisson bracket is given by
\begin{equation}
\label{eq2704-9}
\{q_i,p_j\}=\delta_{ij},\quad\{p_1,p_2\}=\bar S(q),\quad\{q_1,q_2\}=0.
\end{equation}
\end{teo}

{\it Proof}

Let us apply the Legendre transformation to the initial system
\eqref{eq2704-7}:
$$
P_i=\frac{\partial L}{\partial\dot q_i},\quad H=\sum\limits_iP_i\dot
q_i-L\Bigr|_{\dot q_i\to P_i};
$$
here,
\begin{equation}
\label{eq2704-10}
\begin{gathered}
\dot q_i=\frac{\partial H}{\partial P_i},\quad\dot P_1=-\frac{\partial H}{\partial q_1}+
\frac{\partial H}{\partial P_2}S,\quad\dot P_2=-\frac{\partial H}{\partial q_2}-
\frac{\partial H}{\partial P_1}S,\\ S=a_1(q)\dot
q_1+a_2(q)\dot q_2+b(q)=A_1(q)P_1+A_2(q)P_2+B(q).
\end{gathered}
\end{equation}
Using the Liouville equation for the density of the invariant measure
$\rho(q)\,dP_1\,dP_2\,dq_1\,dq_2$ of the system \eqref{eq2704-10}, we
find
$$
\dot q_1\left(\frac1\rho\frac{\partial\rho}{\partial
q_1}-A_2(q)\right)+\dot q_2\left(\frac1\rho\frac{\partial\rho}{\partial
q_2}+A_1(q)\right)=0;
$$
since $\rho$ depends only on the coordinates, each of the brackets should
become zero separately:
$$
\frac1\rho\frac{\partial\rho}{\partial q_1}-A_2(q)=0,\quad
\frac1\rho\frac{\partial\rho}{\partial q_2}+A_1(q)=0.
$$
Now we write equation \eqref{eq2704-8} for $\bar S$, taking into
account the relation $\displaystyle\frac1N\frac{\partial\bar L}{\partial
q_i'}=P_i$:
$$
\bar S=\left(NA_1(q)+\frac{\partial N}{\partial q_2}\right)P_1+
\left(NA_2(q)-\frac{\partial N}{\partial q_1}\right)P_2+B(q).
$$
Thus, if we choose $N(q)=\rho(q)$, then $\bar S=B(q)$, and the first
statement of the theorem is proved.

The second statement can be proved with a straightforward verification of the equations and of
the Jacobi identity.

{\small
\begin{rem}
The time substitution, or the reducing multiplier, for multidimensional systems
is closely related to the
invariant measure, but nevertheless, as shown in \cite{b028}, can differ from
it.

Note also that reduction to the Hamiltonian form is useful for application
of the methods of the perturbation theory, introduction of action-angle
variables, analysis of integrability and non-integrability, etc.
\end{rem}}

The Hamiltonian systems with the bracket \eqref{eq2704-9} are used for
description of systems with a generalized potential (for example, motion of
charged particles in a magnetic field) or systems with gyroscopic forces
\cite{b025}. In this case, the closed 2-form $\bar S(q)\,dq_1\wedge
dq_2$ is called the 2-form of gyroscopic forces. Locally, $\Omega$ can be
represented as the exact differential $\Omega=d\omega$,
$\omega=W_1(q)dq_1+W_2(q)dq_2$ and the equations of motions
\eqref{eq2704-8} can be written in the form of the Lagrange--Euler
equations
$$
\begin{gathered}
\frac d{d\tau}\left(\frac{\partial L_W}{\partial q_i'}\right)-
\frac{\partial L_W}{\partial q_i}=0,\\ L_W=\bar L+W(q,q'),\quad W(q,q')=W_1(q)q_1'+W_2(q)q_2',
\end{gathered}
$$
here the Poisson brackets for the new
momenta $\widetilde p_i=p_i+W_i(q)$ and coordinates $q_i$ are canonical
(i.\,e. $\{q_{i}, \tilde{p}_j\}=\delta_{ij}$).

If the manifold $\cal M$, on which the coordinates $q_1$, $q_2$ are
defined is compact, then the criterion for $\Omega$ to be exact is
$$
\int\limits_{\cal M}\Omega=0.
$$
Thus, if $\int\Omega\neq0$  (i.\,e. the 2-form is not exact) then the
generalized potential $W$ and the corresponding Lagrangian and Hamiltonian
functions have singularities (so-called monopoles) \cite{b023}. In this case, sometimes it
is said that the global representation of the equations of
motion in the (canonical) Hamiltonian form is impossible.

\section{Veselova system}

The Veselova system describes the motion of a rigid body with a fixed
point subject to nonholonomic constraint of the form
$(\boldsymbol\omega,\boldsymbol\gamma)=0$, where $\boldsymbol\omega$ and
$\boldsymbol\gamma$ are the body's angular velocity vector
and the unit vector of the space-fixed axis in the frame of reference
fixed to the body. Thus, for the Veselova constraint, the
projection of the angular velocity onto a space-fixed axis is zero. This
constraint is reciprocal to the Suslov constraint \cite{b024}, for which
the projection of the angular velocity onto a body-fixed axis is zero.

In the moving axes fixed to the body, the equations of motion can be
written as follows \cite{b005,b007}:
\begin{equation}
\label{eq2504-1}
{\bf I}\dot{\boldsymbol\omega}={\bf I}\boldsymbol\omega\times\boldsymbol\omega+
\mu\boldsymbol\gamma+\boldsymbol\gamma\times\frac{\partial U}{\partial\boldsymbol\gamma},
\quad\dot{\boldsymbol\gamma}=\boldsymbol\gamma\times\boldsymbol\omega,
\end{equation}
where $\mu$ is an undetermined multiplier, ${\bf I}={\rm
diag}\,(I_1,I_2,I_3)$ is the tensor of inertia, $U(\boldsymbol\gamma)$ is
the potential energy. The undetermined multiplier $\mu$ can be found by
differentiating the constraint:
\begin{equation}
\label{eq2504-2}
\mu=\frac{\left({\bf
I}\boldsymbol\omega\times\boldsymbol\omega+\boldsymbol\gamma\times\frac{\partial
U}{\partial\boldsymbol\gamma},{\bf
I}^{-1}\boldsymbol\gamma\right)}{\left(\boldsymbol\gamma,{\bf
I}^{-1}\boldsymbol\gamma\right)}.
\end{equation}
In the general case the equations \eqref{eq2504-1} admit the integral of
energy and the geometric integral
\begin{equation}
H=\frac12({\bf I}\boldsymbol\omega,\boldsymbol\omega)+
U(\boldsymbol\gamma),\quad\boldsymbol\gamma^2=1,
\end{equation}
as well as the invariant measure
$\rho_\omega\,d^3\boldsymbol\omega\,d^3\boldsymbol\gamma$ with density
\begin{equation}
\rho_\omega=\sqrt{\left(\boldsymbol\gamma,{\bf I}^{-1}\boldsymbol\gamma\right)}.
\end{equation}
When $U=0$, there is an additional integral
\begin{equation}
\label{eq2504-5}
F=({\bf I}\boldsymbol\omega,{\bf I}\boldsymbol\omega)-({\bf
I}\boldsymbol\omega,\boldsymbol\gamma)^2=|{\bf
I}\boldsymbol\omega\times\boldsymbol\gamma|^2,
\end{equation}
and hence, the system is integrable according to the Euler--Jacobi theorem
\cite{b005}.

{\small
\begin{rem}
The integral \eqref{eq2504-5} is generalized when the Brun potential is
added \cite{b005,b007}. Some other integrable potentials are given in
\cite{b001,b013}.
\end{rem}}

{\small
\begin{rem}
The system \eqref{eq2504-1}, \eqref{eq2504-2} with the constraint
$(\boldsymbol\omega,\boldsymbol\gamma)=0$ was rediscovered in paper
\cite{b015} almost ten years after \cite{b005,b007}. In \cite{b015}, an
explicit integration was performed using sphero-conical coordinates.
\end{rem}}

{\small
\begin{rem}
The Veselova system and the nonholonomic systems (considered below)
describing the rolling motion of bodies belong to the class of so-called $LR$- and
$L+R$-systems on Lie groups \cite{b001,b007}. Several results
on the existence of invariant measure for such
systems are known. We do not consider here these general results, especially useful
for multidimensional generalizations. Note also that the general methods of
reduction of nonholonomic systems were examined in many papers, see for example~\cite{b019}.
\end{rem}}

{\small
\begin{rem}
Generalization of the Veselova constraint
$(\boldsymbol\omega,\boldsymbol\gamma)=d\neq0$ was considered in
\cite{b009}. Using Chaplygin's method of integration for a dynamically
asymmetric ball with non-zero constant of areas \cite{b006},
the author presented an explicit integration  of the equations.
\end{rem}}

It was shown in \cite{b001} that the Veselova system is the Chaplygin system
\eqref{eq2704-7} with $b(q)=0$ and, therefore, upon the time substitution $N\,dt=d\tau$,
it can be written in the Hamiltonian form,  where the reducing
multiplier is $N=\rho^{-1}_\omega$. Let us show this explicitly using the
local coordinates (namely, the Euler angles $\theta$, $\varphi$, $\psi$)
and then apply the obtained canonical Poisson structure of the cotangent
bundle of the sphere $T^{*}S^2$ to construct an algebraic Poisson bracket of
redundant variables $\boldsymbol\omega$, $\boldsymbol\gamma$. With such an
algebraization of the Poisson structure, one can naturally establish
an isomorphism with the Neumann system describing the dynamics of a point on
a sphere in a quadratic potential. This isomorphism was straightforwardly established
in \cite{b005,b007}. Later, we will see that this analogy
can be directly extended to the Chaplygin ball and the general Clebsch
system (which includes the Neumann system as a particular case).

In terms of the Euler angles, the body's angular velocity
$\boldsymbol\omega$ and the unit vector $\boldsymbol\gamma$ are given by
\begin{equation}
\label{eq2904-1}
\boldsymbol\omega=(\dot\psi\sin\theta\sin\varphi+\dot\theta\cos\varphi,
\dot\psi\sin\theta\cos\varphi-\dot\theta\sin\varphi,\dot\psi\cos\theta+\dot\varphi),\quad
\boldsymbol\gamma=(\sin\theta\sin\varphi,\sin\theta\cos\varphi,\cos\theta).
\end{equation}

The equation of the constraint is
\begin{equation}
\label{eq2504-6}
f=(\boldsymbol\omega,\boldsymbol\gamma)=\dot\psi+\cos\theta\dot\varphi=0,
\end{equation}
Eliminating the undetermined Lagrange multiplier from the equations of
motion we can write the equation for $\theta$, $\varphi$ as a Chaplygin system:
\begin{equation}
\label{eq2504-7}
\begin{gathered}
\frac d{dt}\left(\frac{\partial T}{\partial\dot\theta}\right)-
\frac{\partial T}{\partial\theta}+\frac{\partial U}{\partial\theta}=\dot\varphi S,\quad
\frac d{dt}\left(\frac{\partial T}{\partial\dot\varphi}\right)-
\frac{\partial T}{\partial\varphi}+\frac{\partial U}{\partial\varphi}=-\dot\theta S,\\
S=\left.\frac{\partial T_0}{\partial\dot\psi}\right|_{\dot\psi=-\cos\theta\dot\varphi}=
\sin^2\theta\left(\dot\theta(I_2-I_1)\sin\varphi\cos\varphi-\dot\varphi(I_1\cos^2\varphi+
I_2\sin^2\varphi+I_3)\right),
\end{gathered}
\end{equation}
where $U$ is the potential energy of the body in an external field,
$T_0=\frac12(\boldsymbol\omega,{\bf I}\boldsymbol\omega)$ is the kinetic
energy without the constraint, while $T$ is the kinetic energy from which
$\dot\psi$ is eliminated using the constraint
\begin{multline}
\label{eq2504-8}
T=T_0\Bigr|_{\dot\psi=-\cos\theta\dot\varphi}=\frac12I_1(\dot\theta\cos\varphi-
\dot\varphi\sin\varphi\sin\theta\cos\theta)^2+\\
+\frac12I_2(\dot\theta\sin\varphi+\dot\varphi\cos\varphi\sin\theta\cos\theta)^2+
\frac12I_3\dot\varphi^2\sin^4\theta.
\end{multline}

{\small
\begin{rem}
The representation \eqref{eq2504-7} is obtained upon
differentiation under the condition \eqref{eq2504-6}:
$$
\frac{\partial T}{\partial\dot\theta}=\frac{\partial T_0}{\partial\dot\theta},\quad\frac{\partial T}{\partial\dot\varphi}=\frac{\partial T_0}{\partial\dot\varphi}-\cos\theta\frac{\partial T_0}{\partial\dot\psi},\quad\frac{\partial T}{\partial\theta}=\frac{\partial T_0}{\partial\theta}+\dot\varphi\sin\theta\frac{\partial T_0}{\partial\dot\psi},\quad\frac{\partial T}{d\varphi}=\frac{\partial T_0}{\partial\varphi}.
$$
\end{rem}}

\begin{teo}[{\cite{b001}}]
After the time substitution $N\,dt=d\tau$, $N=(\boldsymbol\gamma,{\bf
I}\boldsymbol\gamma)^{-1/2}$, the equations of motion of the Veselova
system take the form of the Euler--Lagrange equations:
\begin{equation}
\label{eq2504-9}
\frac d{d\tau}\left(\frac{\partial L}{\partial\theta'}\right)-
\frac{\partial L}{\partial\theta}=0,\quad\frac d{d\tau}\left(\frac{\partial L}
{\partial\varphi'}\right)-\frac{\partial L}{\partial\varphi}=0,
\end{equation}
where $L=T-U\Bigr|_{\dot\theta=N\theta',\;\dot\varphi=N\varphi'}$ is the
Lagrangian function; after time
substitution it can be written in the form\/{\rm:}
$$
L=\frac12\frac{\left(\boldsymbol\gamma'\times\boldsymbol\gamma,{\bf
I}(\boldsymbol\gamma'\times\boldsymbol\gamma)\right)}
{(\boldsymbol\gamma,{\bf I}^{-1}\boldsymbol\gamma)}-U(\boldsymbol\gamma).
$$
\end{teo}

{\it Proof}

The {\it proof}  is based on a simple computation test: after the time substitution,
the right-hand side $\bar S$ of \eqref{eq2504-9}, calculated by virtue of
\eqref{eq2704-8}, should vanish.
\bigskip

The canonical Hamiltonian form of the equations of motion \eqref{eq2504-9}
can be obtained using the Legendre transformation
\begin{equation}
\label{eq2504-10}
\begin{gathered}
p_\theta=\frac{\partial L}{\partial\theta'},\quad
p_\varphi=\frac{\partial L}{\partial\varphi'},\quad H=p_\theta\theta'+p_\varphi\varphi'-L,\\
\frac{d\theta}{d\tau}=\frac{\partial H}{\partial p_\theta},\quad
\frac{d\varphi}{d\tau}=\frac{\partial H}{\partial p_\varphi},\quad
\frac{dp_\theta}{d\tau}=-\frac{\partial H}{\partial\theta},\quad
\frac{dp_\varphi}{d\tau}=-\frac{\partial H}{\partial\varphi}.
\end{gathered}
\end{equation}

Using the canonical variables of \eqref{eq2504-10} and the time substitution
$(\rho_\omega\sqrt{\det{\bf I}})^{-1}dt=d\tau$, one can write the
equations of motion of the Veselova system  in the Hamiltonian form on
the (co)algebra of the Poisson brackets $e(3)$:
\begin{gather}
\label{eq2504-10-2}
\begin{gathered}
\boldsymbol M=\rho_\omega{\bf I}^{1/2}\boldsymbol\omega,\quad\boldsymbol\Gamma=
\rho^{-1}_\omega{\bf I}^{-1/2}\boldsymbol\gamma,\\
\frac{d\boldsymbol M}{d\tau}=\boldsymbol M\times\frac{\partial H}{\partial\boldsymbol M}+
\boldsymbol\Gamma\times\frac{\partial H}{\partial\boldsymbol\Gamma},\quad
\frac{d\boldsymbol\Gamma}{d\tau}=\boldsymbol\Gamma\times\frac{\partial H}
{\partial\boldsymbol M},
\end{gathered}\\
\label{eq2504-11}
H=\frac12(\boldsymbol\Gamma,{\bf I}\boldsymbol\Gamma)(\boldsymbol M,\boldsymbol M)+
\widetilde U(\boldsymbol\Gamma),
\end{gather}
where $\widetilde U(\boldsymbol\Gamma)=U\left(\rho_\omega{\bf
I}^{1/2}\boldsymbol\Gamma\right)$, and, respectively,
$$
\begin{gathered}
\boldsymbol\gamma^2=\boldsymbol\Gamma^2=1,\quad(\boldsymbol\omega,\boldsymbol\gamma)=
(\boldsymbol M,\boldsymbol\gamma)=0,\quad\rho_\omega=(\boldsymbol\gamma,{\bf I}^{-1}
\boldsymbol\gamma)^{1/2}=(\boldsymbol\Gamma,{\bf I}\boldsymbol\Gamma)^{-1/2},\\
{}\{M_i,M_j\}=\varepsilon_{ijk}M_k,\quad\{M_i,\Gamma_j\}=\varepsilon_{ijk}\Gamma_k,
\quad\{\Gamma_i,\Gamma_j\}=0.
\end{gathered}
$$
Thus, we have a Hamiltonian system with Poisson brackets corresponding to the algebra
$e(3)$, and the four-dimensional symplectic leaf of the structure corresponding to
the real motion is given by $\boldsymbol\gamma^2=1$, $(\boldsymbol
M,\boldsymbol\gamma)=0$ (for the classical Euler--Poisson equations
a similar situation takes place if the constant of areas \cite{b023} is zero). Note also that the
system \eqref{eq2504-10-2}, \eqref{eq2504-11} determines a certain
integrable potential system on a two-dimensional sphere and defines
thereby a certain geodesic flow.

The inverse transformation is
$$
\boldsymbol\omega=(\boldsymbol\Gamma,{\bf I}\boldsymbol\Gamma)^{1/2}{\bf I}^{-1/2}
\boldsymbol M,\quad \boldsymbol\gamma=(\boldsymbol\Gamma,{\bf I}\boldsymbol\Gamma)^{-1/2}
{\bf I}^{1/2}\boldsymbol\Gamma.
$$

Therefore, a search for integrable potentials for the Veselova system is
now reduced to the well-studied problem of search for integrable cases in
a Hamiltonian system on $e(3)$ with Hamiltonian \eqref{eq2504-11}. So, if
$U=0$, then the additional integral \eqref{eq2504-5} can be written as
$$
F=({\bf I}\boldsymbol M,\boldsymbol M)({\bf I}\boldsymbol\Gamma,\boldsymbol\Gamma)-({\bf I}\boldsymbol M,\boldsymbol\Gamma)^2.
$$
(Note that $\{H,F\}=0$ only on the level $(\boldsymbol M,\boldsymbol\gamma)=0$.)

It was noted in \cite{b005,b007} that for $U=0$ the system
\eqref{eq2504-1} is equivalent to the Neumann problem. As we can see, such
equivalence is not a result of the natural
reduction of the Veselova system to the Hamiltonian form
\eqref{eq2504-10-2}, \eqref{eq2504-11} on $e(3)$. It turns out that the
isomorphism with the Neumann system is caused by existence of a
transformation that does not conserve the Poisson brackets but reduces
the vector field to the required form on the level surface $H={\rm
const}$.

Indeed, let us consider a Hamiltonian system on $e(3)$ (in the case $(\boldsymbol
M,\boldsymbol\gamma)=0$) defined by the Hamiltonian
\begin{equation}
\label{eq2504-12}
H=\alpha\frac12\boldsymbol M^2(\boldsymbol\Gamma,{\bf I}\boldsymbol\Gamma)+
\beta\frac12\left((\boldsymbol M,{\bf I}\boldsymbol M)(\boldsymbol\Gamma,{\bf I}
\boldsymbol\Gamma)-(\boldsymbol M,{\bf I}\boldsymbol\Gamma)^2\right).
\end{equation}

It is clear that both terms are the first integrals of the system. The
following holds true:

\begin{pro}\label{pro2704-1}
On a fixed level $\displaystyle\frac{\boldsymbol
M^2(\boldsymbol\Gamma,{\bf I}\boldsymbol\Gamma)}{\det{\bf I}}=c$ and
$(\boldsymbol M, \boldsymbol\Gamma)=0$, the
vector field generated by the Hamiltonian \eqref{eq2504-12} is isomorphic
to the vector field  of the Kirchhoff equations in the Clebsch case with
zero value of constant of areas $(\boldsymbol L,\boldsymbol s)=0$; this
field can be written in the form
\begin{equation}
\label{eq2504-13}
\begin{aligned}
\dot{\boldsymbol s}&=k(\alpha\boldsymbol s\times\boldsymbol L+\beta\boldsymbol s
\times{\bf I}\boldsymbol L),\\
\dot{\boldsymbol L}&=k\left(\alpha c\boldsymbol s\times{\bf I}\boldsymbol s+
\beta(\boldsymbol L\times{\bf I}\boldsymbol L-c(\det{\bf I})\boldsymbol s
\times{\bf I}^{-1}\boldsymbol s)\right),
\end{aligned}
\quad k=-\sqrt{\det{\bf I}}.
\end{equation}
\end{pro}

{\it Proof}

Let us change the variables
$$
\boldsymbol L={\bf I}^{-1/2}\boldsymbol M,\quad\boldsymbol s=(\boldsymbol\Gamma,{\bf
I}\boldsymbol\Gamma)^{-1/2}{\bf I}^{1/2}\boldsymbol\Gamma,
$$
so that the relations $\boldsymbol s^2=\boldsymbol\Gamma^2=1$,
$(\boldsymbol M,\boldsymbol\Gamma)=(\boldsymbol s,\boldsymbol L)=0$ hold.
By virtue of linearity we consider the two cases $\alpha=1$, $\beta=0$ and
$\alpha=0$, $\beta=1$ separately. In the first case the
equations of motion in terms of the new variables read
$$
\begin{aligned}
\dot{\boldsymbol s}&=-\sqrt{\det{\bf I}}\left(\boldsymbol s\times\boldsymbol L+
(\boldsymbol s,\boldsymbol L)\frac{(\boldsymbol s\times{\bf I}^{-1}\boldsymbol s)}
{(\boldsymbol s,{\bf I}^{-1}\boldsymbol s)}\right),\\
\dot{\boldsymbol L}&=-\sqrt{\det{\bf I}}\frac{({\bf I}\boldsymbol L,\boldsymbol L)}
{\det{\bf I}(\boldsymbol s,{\bf I}^{-1}\boldsymbol s)}\boldsymbol s\times{\bf I}\boldsymbol s.
\end{aligned}
$$
Hence, taking into account $(\boldsymbol s,\boldsymbol L)=0$,
$\displaystyle\frac{({\bf I}\boldsymbol L,\boldsymbol L)}{(\boldsymbol
s,{\bf I}^{-1}\boldsymbol s)}=\boldsymbol M^2(\boldsymbol\Gamma,{\bf
I}\boldsymbol\Gamma)=c\det{\bf I}$, we get the required result.

The case $\alpha=0$, $\beta=1$ can be considered analogously.
\bigskip

If we consider $c$ as a constant parameter, then the vector field
\eqref{eq2504-13} is generated on $e(3)$ by the following Hamiltonian
$$
H=k\alpha\left(\frac12\boldsymbol M^2+\frac c2(\boldsymbol\Gamma,{\bf I}\boldsymbol\Gamma)\right)+
k\beta\left(\frac12(\boldsymbol M,{\bf I}\boldsymbol M)-\frac c2\det{\bf I}(\boldsymbol\Gamma,{\bf
I}\boldsymbol\Gamma)\right).
$$
If $\alpha=1$, $\beta=0$, we obtain the Hamiltonian of the Neumann case,
while at $\alpha=0$, $\beta=1$ this is the Hamiltonian of the Brun
problem. With arbitrary $\alpha$, $\beta$, this Hamiltonian corresponds to
the general Clebsch case in the Kirchhoff equations \cite{b012,b023}.
Using the representation \eqref{eq2504-10-2}, \eqref{eq2504-11}, we easily
obtain the following theorem for the Veselova system:

\begin{teo}[{\cite{b005,b007}}]
After time substitution, the vector field of the Veselova problem $($with $U=0)$ on the fixed level
of the integral of energy $H=h={\rm const}$ becomes isomorphic to the vector field of the Neumann problem.
\end{teo}

\section{Chaplygin's ball}

Consider the problem of rolling without sliding of a balanced, dynamically asymmetric ball
on a horizontal plane in the axisymmetric potential force field (we assume
that the geometrical center and the center of mass coincide). We fix a moving
frame of reference to the body and write equations of motion in the
following form \cite{b006}:
\begin{equation}
\label{eq2604-1}
\begin{gathered}
\dot{\boldsymbol M}=\boldsymbol M\times\boldsymbol\omega+\boldsymbol\gamma\times
\frac{\partial U}{\partial\boldsymbol\gamma},\quad\dot{\boldsymbol\gamma}=\boldsymbol\gamma
\times\boldsymbol\omega,\\
\boldsymbol M={\bf I}\boldsymbol\omega+D\boldsymbol\gamma\times(\boldsymbol\omega\times
\boldsymbol\gamma)={\bf I}_Q\boldsymbol\omega,\quad D=mR^2,
\end{gathered}
\end{equation}
where $\boldsymbol\omega$ is the ball's angular velocity,
$\boldsymbol\gamma$ is the vertical unit vector in the moving frame of
reference, ${\bf I}={\rm diag}\,(I_1,I_2,I_3)$ is the ball's tensor of
inertia with respect to its center, $m$ and $R$ are the ball's mass and
radius, and $U=U(\boldsymbol\gamma)$ is the potential of the external
axisymmetric field. The vector $\boldsymbol M$ is the ball's angular
momentum with respect to the point of contact. We present the tensor ${\bf
I}_Q$ in the form
$$
{\bf I}_Q={\bf J}-D\boldsymbol\gamma\otimes\boldsymbol\gamma,\quad {\bf J}={\bf I}+D{\bf E}.
$$

Equations \eqref{eq2504-1} (with an arbitrary potential) admit the
integral of energy, the geometric integral and the integral of areas:
\begin{equation}
\label{eq2604-2} H=\frac12(\boldsymbol
M,\boldsymbol\omega)+U(\boldsymbol\gamma),\quad
(\boldsymbol\gamma,\boldsymbol\gamma)=1,\quad(\boldsymbol
M,\boldsymbol\gamma)=c={\rm const}.
\end{equation}
They also admit the invariant measure indicated by Chaplygin
\cite{b006}, $\rho_\mu\,d^3\boldsymbol M\,d^3\boldsymbol\gamma$, with
density
\begin{equation}
\label{eq2604-3}
\rho_\mu=(\det{\bf I}_Q)^{-1/2}=\left[\det{\bf
J}\left(1-D\left(\boldsymbol\gamma,{\bf J}^{-1}\boldsymbol\gamma\right)
\right)\right]^{-1/2}.
\end{equation}

If there is no external field $(U=0)$, the system
\eqref{eq2504-1} has an additional integral
\begin{equation}
\label{eq2604-4} F=(\boldsymbol M,\boldsymbol M),
\end{equation}
hence, it is integrable according to the Euler--Jacobi theorem
\cite{b006}. In \cite{b006}, the solution of \eqref{eq2604-1} was given in
terms of hyperelliptic functions.

{\small
\begin{rem}
The integral \eqref{eq2604-4} can be generalized to the cases of the
Brun field $U(\boldsymbol\gamma)=\frac k2(\boldsymbol\gamma,{\bf
I}\boldsymbol\gamma)$ \cite{b010} and gyrostat \cite{b025}. Other
integrable potentials (with the zero constant of areas $(\boldsymbol
M,\boldsymbol\gamma)=0$) can be found using the representation of
the system on algebra $e(3)$, which is given below.
\end{rem}

}

{\small {\bf Comment}. Although Chaplygin developed the general method of
reducing multiplier \cite{b003}, he did not apply it to the system \eqref{eq2604-1}.
The paper \cite{b010} indicates possible obstructions to the application of
this method to the system \eqref{eq2604-1}. On the other hand, already
in \cite{b006} the following question was formulated: if the system \eqref{eq2604-1}
can be represented in Hamiltonian form? The problem of
Hamiltonization of Chaplygin's ball was formulated more strictly by
Kozlov \cite{b012} and Duistermaat \cite{b011,b017}. In \cite{b008},
the authors showed numerically that without a time substitution, the
equations of motion of the Chaplygin ball are not Hamiltonian because
the periods of motion for the orbits lying on the two-dimensional invariant resonance
tori are not the same. However, after an appropriate time substitution
the system \eqref{eq2604-1} becomes Hamiltonian and Poisson brackets are
found explicitly (see Ref.~\cite{b002}).

Unfortunately, the authors
of the review \cite{b017} did not succeed in an explicit verifying our
result (perhaps, due to certain misprints in \cite{b002}). Here,
we will prove the result of \cite{b002} using another method and then reveal
an interesting {\it isomorphism between the Chaplygin ball and the Clebsch case} in
the Kirchhoff equations. Another isomorphism has been indicated in
\cite{kf}}.

\smallskip

To describe the ball's rotation let us add
to the system \eqref{eq2604-1} the equations for the remaining direction
cosines:
$$
\dot{\boldsymbol\alpha}=\boldsymbol\alpha\times\boldsymbol\omega,\quad
\dot{\boldsymbol\beta}=\boldsymbol\beta\times\boldsymbol\omega.
$$
Such a system admits two additional integrals linear in velocities
$$
(\boldsymbol M,\boldsymbol\alpha)={\rm const},\quad
(\boldsymbol M,\boldsymbol\beta)={\rm const}.
$$
Under such an extension, the integral manifolds remain
two-dimensional. Hence, the Chaplygin problem with $U=0$ is
degenerate or, as it is sometimes called, superintegrable. From this
viewpoint, Chaplygin's ball is a nonholonomic analog of the Euler--Poinsot
top, a well-known noncommutative integrable system, and the phase space of
this three-degree-of-freedom Hamiltonian system is foliated into two-dimensional tori (not
three-dimensional according to the Liouville theorem).

It was shown in \cite{b002} that, for an arbitrary potential, after the
time substitution and change of variables
\begin{equation}
\label{eq2604-5} \rho_\mu\,dt=d\tau,\quad\boldsymbol L=\rho_\mu\boldsymbol
M,
\end{equation}
the equations of motion \eqref{eq2504-1} take in the Hamiltonian
form:
$$
\frac{dM_k}{d\tau}=\{H,M_k\},\quad\frac{d\gamma_k}{d\tau}=\{H,\gamma_k\}
$$
with the nonlinear Poisson bracket
\begin{equation}
\label{eq2604-6}
\{L_i,L_j\}=\varepsilon_{ijk}\left(L_k-D(\boldsymbol L,\boldsymbol\gamma)
\rho_\mu^2J_iJ_j\gamma_k\right),\quad \{L_i,\gamma_j\}=
\varepsilon_{ijk}\gamma_k,\quad\{\gamma_i,\gamma_j\}=0,
\end{equation}
and the Hamiltonian is the energy \eqref{eq2504-2}, which can be written
in the form:
\begin{equation}
\label{eq2604-7}
H=\frac{\det{\bf J}}2\left(\left(1-D\left(\boldsymbol\gamma,{\bf J}^{-1}
\boldsymbol\gamma\right)\right)
\left(\boldsymbol L,{\bf J}^{-1}\boldsymbol L\right)+D\left({\bf J}^{-1}
\boldsymbol L,\boldsymbol\gamma\right)^2\right)+U(\boldsymbol\gamma).
\end{equation}

Now we show that the system \eqref{eq2604-1} describing the motion of Chaplygin's ball
is a generalized Chaplygin system \eqref{eq2704-7}, and the bracket
\eqref{eq2604-6} can be obtained using the method of reducing multiplier
(see theorem \ref{teo2904-1}).

As in the Veselova problem, we use now the local coordinates: the
Euler angles $\theta$, $\varphi$, $\psi$ and the Cartesian coordinates of
the ball's center $x$, $y$. In the moving frame of reference aligned with the ball's principal axes,
the angular velocity vector and
the normal to the plane are given by \eqref{eq2904-1}.

The equations of the constraints (corresponding to the no slip condition
at the point of contact) can be written in the form
\begin{equation}
\label{eq001}
f_x=\dot x-R\dot\theta\sin\psi+R\dot\varphi\sin\theta\cos\psi=0,\quad
f_y=\dot y+R\dot\theta\cos\psi+R\dot\varphi\sin\theta\sin\psi=0.
\end{equation}
The equations of motion with  Lagrange multipliers are
follows:
\begin{equation}
\label{eq002}
\begin{gathered}
\frac d{dt}\left(\frac{\partial T_0}{\partial\dot
x}\right)=\lambda_x,\quad\frac d{dt}\left(\frac{\partial T_0}{\partial\dot
y}\right)=\lambda_y,\quad\frac
d{dt}\left(\frac{\partial T_0}{\partial\dot\psi}\right)=0,\\
\frac d{dt}\left(\frac{\partial T_0}{\partial\dot\theta}\right)-
\frac{\partial T_0}{\partial\theta}=\lambda_x\frac{\partial f_x}{\partial\dot\theta}+
\lambda_y\frac{\partial f_y}{\partial\dot\theta},\quad\frac d{dt}
\left(\frac{\partial T_0}{\partial\dot\varphi}\right)-
\frac{\partial T_0}{\partial\varphi}=\lambda_x\frac{\partial f_x}{\partial\dot\varphi}+
\lambda_y\frac{\partial f_y}{\partial\dot\varphi},
\end{gathered}
\end{equation}
where $T_0$ is the ball's kinetic energy without taking into account the
constraints \eqref{eq001} (obviously, this energy does not depend on $x$,
$y$, and $\psi$):
$$
T_0=\frac12m(\dot x^2+\dot y^2)+\frac12(\boldsymbol\omega,{\bf I}\boldsymbol\omega).
$$
Eliminating the undetermined multipliers $\lambda_x$ and $\lambda_y$
with the help of the first two equations in \eqref{eq002} and the constraints
\eqref{eq001} we get
$$
\begin{aligned}
\lambda_x\frac{\partial f_x}{\partial\dot\theta}+\lambda_y\frac{\partial f_y}
{\partial\dot\theta}&=-mR^2(\ddot\theta+\dot\psi\dot\varphi\sin\theta),\\
\lambda_x\frac{\partial f_x}{\partial\dot\varphi}+\lambda_y\frac{\partial f_y}
{\partial\dot\varphi}&=-mR^2(\ddot\varphi\sin\theta+\dot\theta\dot\varphi\cos\theta-
\dot\theta\dot\psi)\sin\theta.
\end{aligned}
$$
Hence, the equations of motion for the angles $\theta$ and $\varphi$ do
not depend on $\psi$, but only on $\dot\psi$. Therefore,
$\psi$ is a cyclic variable and can be eliminated using the Routh
reduction procedure; after that the equations of motion for $\theta$ and
$\varphi$ can be written as
\begin{equation}
\label{eq003}
\begin{gathered}
\frac d{dt}\left(\frac{\partial{\cal R}}{\partial\dot\theta}\right)-
\frac{\partial{\cal R}}{\partial\theta}=-\dot\varphi S,\quad \frac d{dt}
\left(\frac{\partial{\cal R}}{\partial\dot\varphi}\right)-\frac{\partial{\cal R}}
{\partial\varphi}=\dot\theta S,\\
S=mR^2\sin\theta\left(\dot\varphi\cos\theta+\dot\psi\right).
\end{gathered}
\end{equation}
Here, ${\cal R}$ is the Routh function:
$$
{\cal R}(\theta,\varphi,\dot\theta,\dot\varphi)=T_0-\dot\psi\frac{\partial T_0}
{\partial\dot\psi},
$$
from which $\dot x$ and $\dot y$ should be eliminated using the equations of
constraints, while $\dot\psi$ is eliminated using the equation for the
cyclic integral,
\begin{multline}
\label{eq004}
\frac{\partial T_0}{\partial\dot\psi}=(I_1-I_2)\dot\theta\sin\theta\sin\varphi\cos\varphi+
I_3\dot\varphi\cos\theta+\\
+\left((I_1\sin^2\varphi+I_2\cos^2\varphi)\sin^2\theta+I_3\cos^2\theta\right)\dot\psi=c=
{\rm const}.
\end{multline}

So, we get the equations of motion in the form of a generalized
Chaplygin system \eqref{eq2704-7}, and, since the system has an invariant measure,
it is possible to write equations \eqref{eq003} in the Hamiltonian
form with the bracket \eqref{eq2704-9}.

Perform the time substitution of the form
\begin{equation}
\label{eq005}
N(\theta,\varphi)dt=d\tau,
\end{equation}
where $N=\rho_\mu$ is the density of the invariant measure
\eqref{eq2604-3}.

According to \eqref{eq2704-8}, the equations of motion, in terms of the new
time, are
\begin{equation}
\label{eq007}
\begin{gathered}
\frac d{d\tau}\left(\frac{\partial\bar{\cal R}}{\partial\theta'}\right)-\frac{\partial\bar{\cal R}}{\partial\theta}=-\varphi'\bar S,\quad \frac d{d\tau}\left(\frac{\partial\bar{\cal R}}{\partial\varphi'}\right)-\frac{\partial\bar{\cal R}}{\partial\varphi}=\theta'\bar S,\\
\bar S=c(I_3+mR^2)mR^2N^3\sin\theta(I_1\cos^2\varphi+I_2\sin^2\varphi+mR^2),
\end{gathered}
\end{equation}
where $\theta'=\displaystyle\frac{d\theta}{d\tau}=N^{-1}\dot\theta$,
$\varphi'=\displaystyle\frac{d\varphi}{d\tau}=N^{-1}\dot\varphi$,
$\bar{\cal R}={\cal R}(\theta,\dot\theta,\varphi,\dot\varphi)
\bigr|_{\dot\theta=N\theta',\;\dot\varphi=N\varphi'}$.

By applying the Legendre transformation to the system \eqref{eq007} we
arrive at

\begin{teo}
Upon the time substitution $\rho_{\mu}dt=d\tau$
the equations of motion \eqref{eq003} for Chaplygin's ball
can be written in the Hamiltonian form\/{\rm:}
$$
\frac{d\theta}{d\tau}=\frac{\partial H}{\partial
p_\theta},\quad\frac{dp_\theta}{d\tau}=-\frac{\partial
H}{\partial\theta}-\bar S\frac{\partial H}{\partial
p_\varphi},\quad\frac{d\varphi}{d\tau}=\frac{\partial H}{\partial
p_\varphi},\quad\frac{dp_\varphi}{d\tau}=-\frac{\partial
H}{\partial\varphi}+\bar S\frac{\partial H}{\partial p_\theta},
$$
with the Poisson bracket of the form
\begin{equation}
\label{eq2704-1}
\{\theta,p_\theta\}=\{\varphi,p_\varphi\}=1,\quad\{p_\varphi,p_\theta\}=\bar
S(\theta,\varphi),\quad\{\theta,\varphi\}=0,
\end{equation}
where
$$
\begin{gathered}
p_\theta=\frac{\partial\bar{\cal R}}{\partial\theta'},\quad
p_\varphi=\frac{\partial\bar{\cal R}}{\partial\varphi'},\\
H=\theta'p_\theta+\varphi'p_\varphi-\bar{\cal R}=\\
=\frac12p_\theta^2\left(I_3\widetilde I_{12}-
D(\boldsymbol\gamma,{\bf I}\boldsymbol\gamma)\right)+
+\frac12\frac{p_\varphi^2}{\sin^2\theta}\left(I_1I_2\sin^2\theta+
I_3\widetilde I_{12}\cos^2\theta-D(\boldsymbol\gamma,{\bf I}\boldsymbol\gamma)\right)+\\
+\frac{p_\theta p_\varphi}{\sin\theta}I_3(I_1-I_2)\cos\theta\sin\theta\sin\varphi\cos\varphi
-\frac{Ncp_\theta}{\sin\theta}(I_1-I_2)(I_3+D\sin^2\theta)\sin\varphi\cos\varphi-\\
-\frac{Ncp_\varphi}{\sin^2\theta}I_3(\widetilde I_{21}+D)+\frac{N^2c^2}{\sin^2\theta}
(I_3+D\sin^2\theta)(\widetilde I_{21}+D),\\
\widetilde I_{12}=I_1\sin^2\varphi+I_2\cos^2\varphi,\quad\widetilde I_{21}=I_1\cos^2\varphi+
I_2\sin^2\varphi.
\end{gathered}
$$
\end{teo}

Expressing the variables $\boldsymbol L=\rho_\mu\boldsymbol M$
\eqref{eq2604-5} in terms of the local variables $\theta$, $\varphi$,
$p_\theta$, and $p_\varphi$, we find
$$
L_1=p_\theta\cos\varphi-p_\varphi\frac{\cos\theta\sin\varphi}{\sin\theta}+
cN\frac{\sin\varphi}{\sin\theta},\quad L_2=-p_\theta\sin\varphi-p_\varphi
\frac{\cos\theta\cos\varphi}{\sin\theta}+cN\frac{\cos\varphi}{\sin\theta},\quad L_3=p_\varphi.
$$
Using such a transformation one can straightforwardly obtain \eqref{eq2604-6}.

Let us consider in more detail the integrable case $U=0$ with zero
constant of areas $(\boldsymbol M,\boldsymbol\gamma)=(\boldsymbol
L,\boldsymbol\gamma)=0$, because the bracket \eqref{eq2604-6} in this case
corresponds to the algebra $e(3)$. We write the Hamiltonian
\eqref{eq2604-2} (omitting unessential multipliers) and the additional
integral \eqref{eq2604-4} in terms of the variables $\boldsymbol L$,
$\boldsymbol\gamma$ as follows
$$
\begin{gathered}
H=\frac12\boldsymbol L^2(\boldsymbol\gamma,{\bf B}\boldsymbol\gamma)-
\frac12\left[(\boldsymbol L,{\bf B}\boldsymbol L)(\boldsymbol\gamma,{\bf B}
\boldsymbol\gamma)-(\boldsymbol\gamma,{\bf B}\boldsymbol L)^2\right],\\
F=\boldsymbol L^2(\boldsymbol\gamma,{\bf B}\boldsymbol\gamma),\quad{\bf B}=1-D{\bf J}^{-1}=
{\bf I}{\bf J}^{-1}.
\end{gathered}
$$
Using proposition \ref{pro2704-1} from the previous section, we get
the following result:

\begin{teo}
On a fixed level $(\boldsymbol M,\boldsymbol M)={\rm const}$ and $(\boldsymbol
M,\boldsymbol\gamma)=0$ the vector field \eqref{eq2604-1}, with $U=0$,
after the time substitution \eqref{eq2604-5} and
the change of variables
$$
\boldsymbol s=(\boldsymbol\gamma,{\bf B}\boldsymbol\gamma)^{-1/2}{\bf B}^{-1/2}
\boldsymbol\gamma,\quad \widetilde{\boldsymbol L}={\bf B}^{-1/2}\boldsymbol L
$$
is reduced to the vector field of the Clebsch case in the Kirchhoff
equations with zero constant of areas.
\end{teo}



\section{Realization of constraints. Chaplygin's ball with the Veselova constraint}

In the paper by A.\,P.\,Veselov and L.\,Ye.\,Veselova \cite{b007}, as well
as in \cite{b017}, the authors consider the problem of rolling on a
plane of a balanced, dynamically asymmetric ball (Chaplygin's ball) with an
additional nonholonomic constraint (the Veselova constraint):
$$
(\boldsymbol\omega,\boldsymbol E)=0,
$$
where $\boldsymbol E$ is the unit vector of a space-fixed axis.

\begin{figure}[!ht]
\centering
\includegraphics{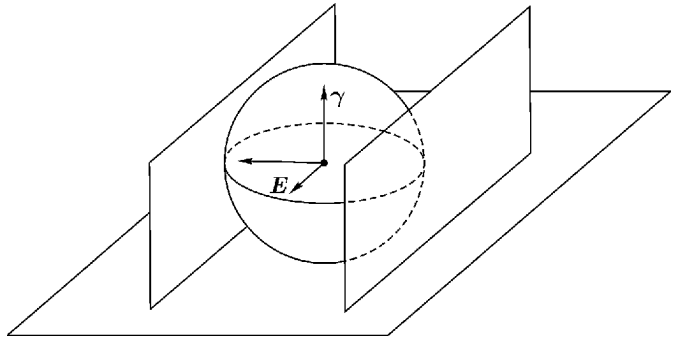}
\caption{A ball rolling along a straight line (A.\,P.\,Veselov,
L.\,Ye.\,Veselova{)}.} \label{ves-01.eps}
\end{figure}

The case $\boldsymbol E\perp\boldsymbol\gamma$, with $\boldsymbol\gamma$
being normal to the plane of contact, was considered in \cite{b007} where
the system was proved to be integrable according to the Euler--Jacobi
theorem. Besides, a realization of this constraint
by means of absolutely smooth walls was offered in
\cite{b007} (see Fig.
\ref{ves-01.eps}) so that the ball moves along the straight line on the plane.

In \cite{b017}, the case $\boldsymbol E\parallel\boldsymbol\gamma$ was
considered and it was supposed that this system should describe the motion of
a (massive) rubber ball on a plane (an ordinary Chaplygin's ball,
considered above, was called in this paper a marble ball). Later, we will
show that this system is equivalent to the Veselova system \eqref{eq2504-1} (and thereby
answer the question of the
integrability of a rubber ball's motion on a plane, formulated in \cite{b017}).

Here, we consider the general situation, assuming that $\boldsymbol E$
is an arbitrary space-fixed unit vector. We start with describing a possible
realization of such a system (or constraints), since it is obvious that the
ball-upon-a-plane realization in this case is impossible.
Basing on results of \cite{b020,b018,b014,b016}, we
offer a more general realization for compositions of Chaplygin
constraints, Suslov constraints and Veselova constraints, using only perfect
rolling (without dissipation of energy).

\begin{figure}[!ht]
\centering\includegraphics{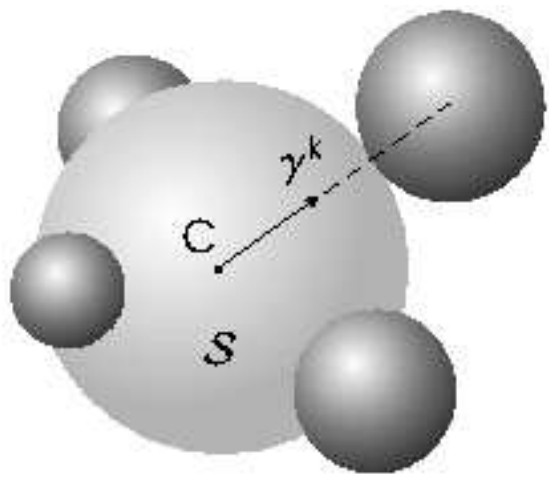}
\smallskip
\caption{The spherical support (Yu.\,N.\,Fedorov)} \label{fed-1.eps}
\end{figure}

Let us start with a {\em spherical support} \cite{b020}. This system
describes the motion of a rigid body with a fixed point $O$ which is
enclosed in a spherical shell; the shell touches an arbitrary number of massive
dynamically symmetric balls with fixed centers (Fig. \ref{fed-1.eps}). It is supposed
that there is no sliding at the points of contact of the balls and
the shell. As was shown in
\cite{b020} this system is integrable with an arbitrary number of
balls. In the case of a single external ball, we have a problem
equivalent to the problem of Chaplygin's ball.

\begin{figure}[ht]
\centering\includegraphics{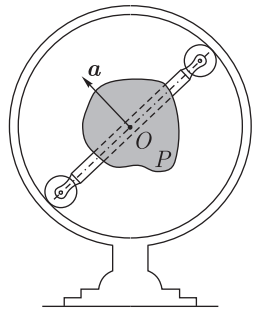}
\smallskip
\caption{Wagner's realization of the Suslov constraint} \label{ves-05.eps}
\end{figure}

\begin{figure}[!ht]
\centering\includegraphics{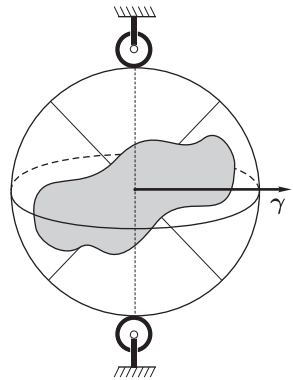}
\smallskip
\caption{Realization of the Veselova constraint} \label{fig2904-1}
\end{figure}

\begin{figure}[!ht]
\centering\includegraphics{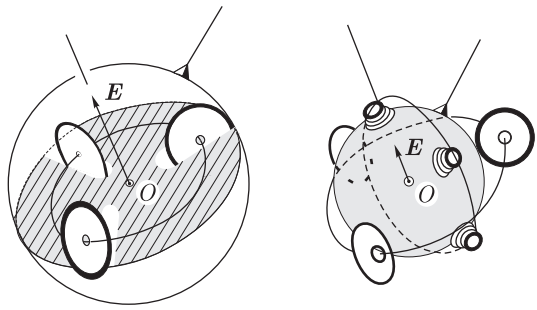}
\smallskip
\caption{The nonholonomic joint} \label{fig2904-2}
\end{figure}

Another system, which can be used for realization of nonholonomic
constraints, is called a {\it nonholonomic joint} \cite{b014}. The
original version of this construction was offered in \cite{b018} as a
realization of the constraint of the Suslov problem
$(\boldsymbol\omega,\boldsymbol a)=0$, where $\boldsymbol a$ is a
body-fixed vector. In this case, flat wheels
(disks) are attached to the body with the fixed point; these wheels roll
without sliding upon the interior surface of a fixed spherical shell
(see Fig. \ref{ves-05.eps}). It
is supposed that a wheel is so sharp that its velocity in the direction
perpendicular to its plane is zero. Similarly, one can consider the motion
of a body with a fixed point when the body is enclosed in a spherical shell,
which is touched by wheels (disks), whose axes are fixed in space (Fig.
\ref{fig2904-1}). It is clear that in the simplest case, we obtain the
constraint of the Veselova problem
$(\boldsymbol\omega,\boldsymbol\gamma)=0$, where $\boldsymbol\gamma$ is
the vector lying in the disk's plane (this result was mentioned in
\cite{b016}). In \cite{b014}, a similar realization of a similar
constraint was offered: a frame with wheels (disks) is attached to the
interior or exterior surface of a spherical shell (Fig. \ref{fig2904-2}).
Such a construction ensures equality of projections of the angular
velocities $\boldsymbol\omega_s$ of the spherical shell and
$\boldsymbol\omega_f$ of the frame with wheels (and, correspondingly, of
the bodies attached to them) onto the axis $\boldsymbol E$ which is
orthogonal to the plane containing the disks' axes:
$$
(\boldsymbol\omega_s,\boldsymbol E)=(\boldsymbol\omega_f,\boldsymbol E).
$$
If the frame with wheels is fixed in space, then we have the Veselova
constraint, while the space-fixed spherical shell gives the Suslov
constraint.

\begin{figure}[!ht]
\centering\includegraphics{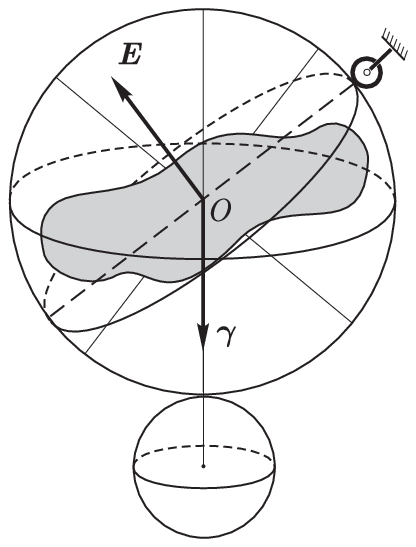}
\smallskip
\caption{}
\label{fig2904-3}
\end{figure}

Now let us consider a combination of a spherical support and a
nonholonomic joint, where the body with a fixed point is enclosed in a
spherical shell which is in contact with a single ball and a single disk (see Fig.
\ref{fig2904-3}).

In the frame of reference aligned with the principal axes of the body, the
equations of constraints read
\begin{equation}
\label{eq2704-2}
R\boldsymbol\omega\times\boldsymbol\gamma+R_1\boldsymbol\omega_1\times\boldsymbol\gamma=0,
\quad(\boldsymbol\omega,\boldsymbol E)=0,
\end{equation}
where $\boldsymbol\omega$ is the body's angular velocity, $R$ is the
radius of the spherical shell, $\boldsymbol\omega_1$ and $R_1$ are the
angular velocity and radius of the adjoining ball, $\boldsymbol\gamma$ is
the unit vector of the axis through the balls' centers, and
$\boldsymbol E$ is the normal vector to the plane that contains the ball's
center and the disk's axis.

The equations of motion with undetermined multipliers look as follows:
\begin{equation}
\label{eq2704-3}
\begin{gathered}
{\bf I}\dot{\boldsymbol\omega}={\bf I}\boldsymbol\omega\times\boldsymbol\omega+
R\boldsymbol\gamma\times\boldsymbol N+\mu\boldsymbol E+\boldsymbol M_Q,\quad
D_1\dot{\boldsymbol\omega}_1=D_1\boldsymbol\omega_1\times\boldsymbol\omega+
R_1\boldsymbol\gamma\times\boldsymbol N,\\
\dot{\boldsymbol\gamma}=\boldsymbol\gamma\times\boldsymbol\omega,\quad\dot{\boldsymbol E}=
\boldsymbol E\times\boldsymbol\omega,
\end{gathered}
\end{equation}
where ${\bf I}={\rm diag}\,(I_1,I_2,I_3)$ is the body's tensor of inertia,
$D_1$ is the scalar tensor of inertia of the adjoining ball, $\boldsymbol
N=(N_1,N_2,N_3)$ and $\mu$ are the undetermined multipliers that
correspond to reaction of the constraints \eqref{eq2704-2}, and
$\boldsymbol M_Q$ is the moment of the external forces. Using the second
equation of \eqref{eq2704-3} we find that
$(\boldsymbol\omega_1,\boldsymbol\gamma)^{\text{{\Large$\cdot$}}}=0$,
therefore,
$$
(\dot{\boldsymbol\omega}_1,\boldsymbol\gamma)=-(\boldsymbol\omega_1,\dot{\boldsymbol\gamma})=-(\boldsymbol\omega_1,\boldsymbol\gamma\times\boldsymbol\omega).
$$
Using this relation and the second equation in \eqref{eq2704-3}, we
eliminate $\boldsymbol\gamma\times\boldsymbol N$ from the remaining
equations. As a result, we obtain
\begin{equation}
\label{eq2704-4}
{\bf I}\dot{\boldsymbol\omega}+D\boldsymbol\gamma\times(\dot{\boldsymbol\omega}\times
\boldsymbol\gamma)={\bf I}\boldsymbol\omega\times\boldsymbol\omega+\mu\boldsymbol E+
\boldsymbol M_Q,\quad D=\frac{R^2}{R_1^2}D_1,\quad\dot{\boldsymbol\gamma}=
\boldsymbol\gamma\times\boldsymbol\omega,\quad\dot{\boldsymbol E}=\boldsymbol E\times
\boldsymbol\omega.
\end{equation}
Comparing this with \eqref{eq2504-1} and \eqref{eq2604-1}, we conclude
that these equations coincide with the equations for the rolling of
Chaplygin's ball with the additional Veselova constraint; in this case the direction of
the vector $\boldsymbol E$ can be arbitrary. The undetermined
multiplier $\mu$ can be found from the relation $(\boldsymbol
E,\boldsymbol\omega)^{\text{{\Large$\cdot$}}}=0$:
$$
\mu=-\frac{({\bf I}\boldsymbol\omega\times\boldsymbol\omega+\boldsymbol
M_Q,{\bf I}^{-1}\boldsymbol E)}{(\boldsymbol E,{\bf I}_Q^{-1}\boldsymbol
E)},\quad{\bf I}_Q={\bf
J}-D\boldsymbol\gamma\otimes\boldsymbol\gamma,\quad {\bf J}={\bf I}+D.
$$

By straightforward calculations one can show that if $\boldsymbol M_Q$ does not depend on
$\boldsymbol\omega$, equations \eqref{eq2704-4} have the invariant
measure $\rho_{\omega}\,d^3\boldsymbol\omega\,d^3\boldsymbol\gamma$ with
density
\begin{equation}
\label{eq2704-5}
\rho_{\omega}=\left((\boldsymbol E,{\bf I}_Q^{-1}\boldsymbol E)\det{\bf I}_Q\right)^{1/2}.
\end{equation}
There are also obvious geometric integrals
$$
\boldsymbol\gamma^2=1,\quad\boldsymbol
E^2=1,\quad(\boldsymbol\gamma,\boldsymbol E)={\rm const}.
$$
In the potential force field $\boldsymbol
M_Q=\boldsymbol\gamma\times\displaystyle\frac{\partial
U}{\partial\boldsymbol\gamma}+\boldsymbol E\times\displaystyle\frac{\partial
U}{\partial\boldsymbol E}$ and the energy is also conserved:
$$
H=\frac12({\bf I}_Q\boldsymbol\omega,\boldsymbol\omega)+U(\boldsymbol\gamma,\boldsymbol E),
$$
where $U(\boldsymbol\gamma,\boldsymbol E)$ is the potential energy of the
external forces. If there are no external forces ($U=0$) and ($\boldsymbol
E\times\boldsymbol\gamma\neq0$), then there are two additional
integrals:
\begin{equation}
\label{eq2704-6}
\begin{gathered}
F_1=(\boldsymbol K,\boldsymbol E\times\boldsymbol\gamma),\quad
F_2=\left(\boldsymbol K,\boldsymbol E\times(\boldsymbol E\times\boldsymbol\gamma)\right),\\
\boldsymbol K={\bf I}_Q\boldsymbol\omega-({\bf
I}_Q\boldsymbol\omega,\boldsymbol E) \boldsymbol E;
\end{gathered}
\end{equation}
hence, the system \eqref{eq2704-4} is integrable (according to the
Euler--Jacobi theorem).

In order to prove that the integrals $F_1$, $F_2$ exist, let us write the
equations of evolution of the vector $\boldsymbol K$:
$$
\dot{\boldsymbol K}=\boldsymbol K\times\boldsymbol\omega.
$$
Hence, the vector $\boldsymbol K$ is fixed in space, and all its
projections onto fixed axes are conserved, but since $(\boldsymbol
K,\boldsymbol E)\equiv0$, only two independent integrals remain.

Therefore, the system \eqref{eq2704-4} (when $\boldsymbol
E\times\boldsymbol\gamma\neq0$) is almost identical to the system
considered in \cite{b007} (where an additional constraint $(\boldsymbol E,\boldsymbol\gamma)=0$ is
imposed). Note that the system
\eqref{eq2704-4} with $U=0$ has not yet been integrated in terms of
quadratures.

In the special case $\boldsymbol E=\boldsymbol\gamma$ the integrals
\eqref{eq2704-6} are identically zero, but from the equation of the
constraint $(\boldsymbol\omega,\boldsymbol\gamma)=0$ we find that
$(\boldsymbol\omega,\boldsymbol\gamma)^{\text{{\Large$\cdot$}}}=
(\dot{\boldsymbol\omega},\boldsymbol\gamma)=0$ and, hence,
$$
{\bf I}_Q\boldsymbol\omega={\bf
J}\boldsymbol\omega-D(\boldsymbol\omega,\boldsymbol\gamma)
\boldsymbol\gamma={\bf J}\boldsymbol\omega,\quad{\bf
I}_Q\dot{\boldsymbol\omega}={\bf
J}\boldsymbol\omega-D(\dot{\boldsymbol\omega},
\boldsymbol\gamma)\boldsymbol\gamma={\bf
J}\dot{\boldsymbol\omega},\quad{\bf
I}\boldsymbol\omega\times\boldsymbol\omega= {\bf
J}\boldsymbol\omega\times\boldsymbol\omega.
$$
Therefore, in this case after the change ${\bf I}\to\bf J$
the system \eqref{eq2704-4} is equivalent to the Veselova
system \eqref{eq2504-1}. Thus, we {\it
answer the question of integrability of rolling of a
rubber ball formulated in \cite{b017}}.

\section{Nonholonomic Jacobi problem}

Here we consider another class of nonholonomic systems with an
invariant measure and first integrals; however, it is still unknown, if the systems'
equations of motion can be written in the Hamiltonian form.

Let a dynamically symmetric ball $({\bf I}=D_0{\bf E})$ of radius $R$
roll without sliding upon a fixed surface. In a fixed Cartesian frame of
reference, the surface, on which the ball's
center moves, is given by $\Phi(\boldsymbol x)=0$. (It is clear that the
surface, on which the ball moves, is equidistant to the surface
$\Phi(\boldsymbol x)=0$, but the equations of motion take a simpler form
if the radius-vector of the center $\boldsymbol x$ is used.)

The equations of the constraint and the equations of motion with
undetermined multipliers are
$$
\begin{gathered}
\boldsymbol v-R\boldsymbol\omega\times\boldsymbol n=0,\quad
\boldsymbol n=\frac{\nabla\Phi(\boldsymbol x)}{|\nabla\Phi(\boldsymbol x)|},\\
m\dot{\boldsymbol v}=\boldsymbol N-\frac{\partial U}{\partial\boldsymbol
x},\quad \mu\dot{\boldsymbol\omega}=R\boldsymbol N\times\boldsymbol n,
\end{gathered}
$$
where $\boldsymbol v=\dot{\boldsymbol x}$ is the velocity of the center of
mass, $\boldsymbol\omega$ is the ball's angular velocity, $m$ is the mass
of the ball, $\boldsymbol N$ is the constraint reaction, $\boldsymbol n$
is the normal to the surface, and $U(\boldsymbol x)$ is the potential
energy of the external field. All the vectors are supposed to be projected
onto the fixed system.

\begin{figure}[!ht]
\centering
\includegraphics{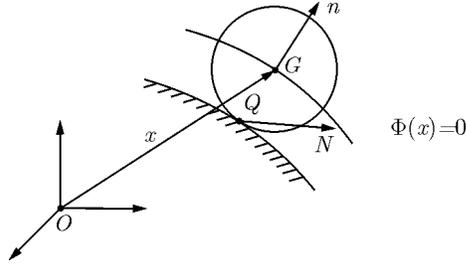}
\smallskip
\caption{Rolling of a ball on a surface ($G$ is the center of mass, $Q$
is the point of contact of the ball and the surface)}
\label{sphris3.eps}
\end{figure}

After eliminating the undetermined multipliers $\boldsymbol N$, we obtain
the equations of motion in the form
\begin{equation}
\label{eq2804-1}
\left(m+\frac{D_0}{R^2}\right)\dot{\boldsymbol\omega}=m(\boldsymbol\omega,\boldsymbol n)
\dot{\boldsymbol n}+R\boldsymbol n\times\frac{\partial U}{\partial\boldsymbol x},\quad
\dot{\boldsymbol x}=R\boldsymbol\omega\times\boldsymbol n,
\end{equation}
where $\dot{\boldsymbol n}$ is expressed, using the equation of the
surface, as follows:
$$
\dot
n_i=\frac1{|\nabla\Phi|}\sum\limits_k\left(\frac{\partial^2\Phi}{\partial
x_i\partial
x_k}-\frac1{|\nabla\Phi|^2}\sum\limits_j\left(\frac{\partial\Phi}{\partial
x_i}\right)\left(\frac{\partial\Phi}{\partial
x_j}\right)\frac{\partial^2\Phi}{\partial x_j\partial x_k}\right)\dot x_k.
$$

Equations \eqref{eq2804-1} admit the geometric integral and the
integral of energy
\begin{equation}
\label{eq2804-2}
\Phi(\boldsymbol x)=0,\quad H=\frac12(D_0+D)\boldsymbol\omega^2-
\frac12\mu(\boldsymbol\omega,\boldsymbol n)^2+U(\boldsymbol x),\quad
D=mR^2.
\end{equation}
Besides, there is also the invariant measure
$\rho_\omega\,d^2\boldsymbol\omega\,d^3\boldsymbol x$ with density
\cite{b026}
\begin{equation}
\label{eq2804-3}
\rho_\omega=|\nabla\Phi(\boldsymbol
x)|=\sqrt{\sum\limits_{k=1}^3\left(\frac{\partial\Phi}{\partial
x_k}\right)^2}.
\end{equation}

{\it Is it possible} ({\it using, say, the method of reducing
multiplier}\/) {\it to represent the equations \eqref{eq2804-1} in the
Hamiltonian form}\/?

Obviously, in the general case the answer to this question is  `no'. Indeed,
writing \eqref{eq2804-1} in terms of the local coordinates on the surface
$\Phi(\boldsymbol x)=0$, we obtain the system of five equations with the
integral of energy. If this system is Hamiltonian, then the corresponding
Poisson bracket is necessarily degenerate (the rank of the Poisson bracket
is always even); consequently, there should exist a Casimir function
$F(\boldsymbol x,\boldsymbol\omega)$ independent of the energy
\eqref{eq2804-2}. But if such a function existed, it would have been
an integral of equations \eqref{eq2804-1}, which, as numerical experiments show, generally does not
exist.

There is an important special case of \eqref{eq2804-1}, when the surface
is a three-axial ellipsoid $\Phi(\boldsymbol x)=(\boldsymbol x,{\bf
B}^{-1}\boldsymbol x)-1=0$, ${\bf B}={\rm diag}\,(b_1,b_2,b_3)$ (more
precisely, an arbitrary quadric surface). This is the so-called {\it
nonholonomic Jacobi problem} \cite{b004}, similar to the problem of
geodesics on an ellipsoid. It is shown in \cite{b004} that in this case
there is an additional integral
\begin{equation}
\label{eq2804-4} K=\frac{(\boldsymbol\omega\times\boldsymbol n,{\bf
B}^{-1}\boldsymbol\omega\times\boldsymbol n)}{(\boldsymbol n,{\bf
B}\boldsymbol n)}.
\end{equation}
The integral \eqref{eq2804-4} is similar to the Joachimstahl integral in the classical
problem of
geodesics and can be extended to the case when the potential
$U(x)=\displaystyle\frac k2\boldsymbol
x^2+\frac12\sum\limits_i\frac{c_i}{x_i^2}$, $k,c_i={\rm const}$, is added.

If the nonholonomic Jacobi problem can be written
in the Hamiltonian form (possibly with time substitution)? It is a very complicated question.
On one hand, as
numerical experiments show, the system in this case behaves chaotically and is
nonintegrable \cite{b004}, so we cannot find here any obstacles for
Hamiltonian representation, typical for integrable systems. Besides, the
two-dimensional Poincar\'e map, which can be constructed on the level
surface of the integrals \eqref{eq2804-2} and \eqref{eq2804-4}, has a
measure and is symplectic; therefore, this map can be generated into the
flow of a Hamiltonian system (see, for example, \cite{b027}). On the other
hand, the method of reducing multiplier and the
explicit Poisson structure fitting do not give the required results.
Note that the
possibility of hamiltonization essentially depends on smoothness,
analiticity, or algebraicity of the sought-for Poisson structure.
Here, we do not consider these
issues.

Another nonholonomic problem (somewhat simpler than the previous one,
because there exists a certain integrable limit problem) concerns the
system considered in \cite{b021}: it is a problem of the {\it spherical
suspension}. It is supposed in this case that a dynamically asymmetric
ball (Chaplygin's ball) rolls on the surface of a fixed sphere. An
analysis of the problem's integrability is given in \cite{b021,b029}.

\section{Acknowledgements}

This work was supported from the program <<State Support for Leading
Scientific Schools>> (grant~136.2003.1); additional support was
provided by the Russian Foundation for Basic Research
(04-05-64367 and
05-01-01058), CRDF (RU-M1-2583-MO-04) and INTAS
(04-80-7297).


\begin{thebibliography}{99}

\bibitem{b001}
Fedorov Yu.\,N., Jovanovi\'c B. {\it Nonholonomic $LR$-Systems as
generalized Chaplygin Systems with an Invariant Measure and Flows on
Homogeneous Spaces}, J. of Nonlinear Science, 2004, v. 1, \No\,14, p.
341--381.

\bibitem{b002}
Borisov A.\,V., Mamaev I.\,S. {\it Chaplygin's ball rolling problem
is Hamiltonian}, Mat. Zametki, 2001, v. 70, \No\,5, p. 793--795 (English transl: Math. Notes, 2001,
v. 70, \No\,5-6, p. 720--723).

\bibitem{b003}
Chaplygin S.\,A. {\it On the theory of motion of nonholonomic systems. Theorems on the reducing
    multiplier}, Mat. Sbornik, 1911, v. 28, \No 2, p. 303--314. (in Russian)

\bibitem{b004}
Borisov A.\,V., Mamaev I.\,S., Kilin A.\,A.
{\it A new integral in the problem of rolling a ball on an arbitrary ellipsoid}, Dokl. RAN,
2002, v. 385, \No\,3, p. 338--341. (in Russian)

\bibitem{b005}
Veselova L.\,E. {\it
New cases of integrability of the equations of motion of a rigid body
in the presence of a nonholonomic constraint}, in:  Geometry,
Dif. Equations and Mechanics. Moscow, MSU press, 1986, p. 64--68. (in Russian)

\bibitem{b006}
Chaplygin S.\,A. {\it On ball's rolling on horizontal plane},
Collection of works. V.\,1. OGIZ, 1948, p. 76--101. (in Russian)

\bibitem{b007}
Veselova A.\,P., Veselova L.\,E.
{\it Integrable nonholonomic systems on Lie groups}, Mat. Zam., 1988, v,.44.
 \No\,5, p. 604--619.

\bibitem{b008}
Borisov A.\,V., Mamaev I.\,S. {\it Obstacle to the reduction of nonholonomic systems
to the Hamiltonian form}, Doklady RAN, 2002, Vol. 387, \No\,6. p.764--766 .

\bibitem{b009}
Fedorov Yu.\,N. {\it On two integrable nonholonomic systems in
classical mechanics}, Vestn. MGU, Ser. mat. mech., 1989, \No\,4, p.
38--41.

\bibitem{b010}
Kozlov V.\,V. 39] {\it On the integrability theory of equations of nonholonomic
mechanics},
Advances in Mechanics, 1985, V. 8, \No 3, p. 85-–107 (Russian); Reg. \& Chaot. Dyn., 2002, v.
7, \No. 2, p. 161--176.

\bibitem{b011}
Duistermaat J.\,J. {\it Chaplygin's sphere}, in Cushman R., Duistermaat
J.\,J., \'Sniatycki J. {\it Chaplygin and the Geometry of Nonholonomically
Constrained Systems} (in preparation), 2000.

\bibitem{b012}
Kozlov V.\,V. {Symmetry, Topology and Resonances in Hamiltonian Mechanics.}
Springer-Verlag, 1996.

\bibitem{b013}
Dragovi\'c V., Gaji\'c B., Jovanovi\'c B. {\it Generalizations of
classical integrable nonholonomic rigid body systems}, J. Phys. A, 1998,
v. 31:49, p. 9861--9869.

\bibitem{b014}
Kharlamov A.\,P., Kharlamov M.\,P. {\it Nonholonomic joint}, Mekh. tverd. tela,
NAS of Ukraine, 1995, v. 27, p. 1--7.

\bibitem{b015}
Kharlamov A.\,P. {\it On the inertial motion of a body with a fixed point subject
to nonholonomic constraint}, Mekh. tverd. tela, NAS of Ukraine, 1995, v. 27, p. 21--31.

\bibitem{b016}
Borisov A.\,V., Mamaev I.\,S.  {\it Chaplygin's ball. The Suslov problem and Veselova's problem.
Integrability and realization of constraints}, in:
Borisov A. V., Mamaev I. S. (Eds) Nonholonomic Dynamical Systems,
Moscow - Izhevsk: Institute of Computer Science, 2002, p. 118-130

\bibitem{b017}
Ehlers K., Koiller J., Montgomery R., Rios P. {\it Nonholonomic systems
via moving frames}\/: {\it certain equivalence and Chaplygin
Hamiltonization}, in {\it The Breadth of Symplectic and Poisson Geometry}
(Eds. J.\,E.\,Marsden, T.\,S.\,Ratiu), Festschrift in honor of Alain
Weinstein, Ser. Progress in Mathematics, v. 232, 2005, Birkh\"auser.

\bibitem{b018}
Vagner V. {\it Geometrical interpretation of the motion
of nonholonomic dynamical systems}, Proceedings of
the seminar on vector and tensor analysis, 1941,
v. 5, p. 301--327. (In Russian.)

\bibitem{b019}
Bloch A.\,M. (with Baillieul J., Crouch P., Marsden J.) {\it Nonholonomic
Mechanics and Control}, Springer, 2000, 483 p.

\bibitem{b020}
Fedorov Yu.\,N. On the motion of a rigid body in a spherical support. Vestn. Moskov.
Univ. Ser. I, Mat. Mekh., 1988, \No\,5, p. 38–41 (Russian).

\bibitem{b021}
Borisov A.\,V.,  Fedorov Yu.\,N. {\it On two modified integrable problems of dynamics},
Vestn. Moskov. Univ. Ser. I, Mat. Mekh., 1995, \No\,6, p. 102--1055 (Russian).

\bibitem{b022}
Cantrijin F., de L\'eon M., de Diego D. {\it On the geometry of
generalized Chaplygin systems}, Math. Proc. Camb. Phil. Soc., 2002, v.
132, p. 323--351.

\bibitem{b023}
Arnold V.\,I., Kozlov V.\,V., Neishtadt A.\,I. {\it  Mathematical aspects of classical
and celestial mechanics}, Itogi Nauki i Tekhniki. Sovr. Probl. Mat. Fundamental’nye
Napravleniya, Vol. 3, VINITI, Moscow 1985. English transl.: Encyclopadia of Math.
Sciences, Vol.3, Springer-Verlag, Berlin 1989.

\bibitem{b024}
Suslov G.\,K. {\it Theoretical mechanic}, Gostekhizdat, Moskva- Leningrad, 1951. (in Russian)

\bibitem{b025}
Markeev A.\,P. {\it On integrability of problem on rolling
of ball with multiply connected cavity filled by ideal
liquid}, Proc. of USSR Acad. of Sciences, Rigid body
mech, 1986, v. 21, \No 1, p. 64--65.


\bibitem{b026}
Yaroschuk V.\,A. {\it New cases of existence of integral
invariant in problem of rolling without sliding of rigid
body on fixed surface}, MGU Bull., Math.\& mech. series.
1992, \No\,6, p. 26--330.

\bibitem{b027}
Douady R. {\it Une d\'emonstration directe de l'\'ecvivalence des
th\'eor\`emes de tores invariants pour diff\'eomorphismes et champs de
vecteures}, C. R. Acad. Sci. Paris, Ser. I, Math., 1982, v. 295(2), p.
201--204.

\bibitem{b028}
Borisov A.\,V., Mamaev I.\,S. {\it Rolling of a rigid body on plane and
sphere. Hierarchy of dynamics}, Reg. \& Chaot. Dyn., 2002, v. 7, \No\,2, p.
177--200.

\bibitem{b029}
Borisov A.\,V.,  Tsygvintsev A.\,V. {\em Kovalevskaya's Method in Rigid Body Dynamics },
Прикл. Мат. Мех., 1997, v.~61, \No\,1, p.~30--36.
(English transl: Appl. Maths Mechs., 1997, v. 61, \No.1, pp. 27-32.

\bibitem{koil}
Koiller J. {\em Reduction of some classical non-holonomic systems with
symmetry.} Arch. Rational. Mech. Anal., 1992, v. 118, p. 113-148.

\bibitem{kf}
Kozlov V.\,V., Fedorov Yu.\,N.
{\em A Memoir on Integrable Systems}, Springer-Verlag, in press.


\end{thebibliography}
\end{document}